\documentclass{article}

\usepackage{arxiv}

\usepackage[utf8]{inputenc} 
\usepackage[T1]{fontenc}    
\usepackage{url}            
\usepackage{booktabs}       
\usepackage{amsfonts}
\usepackage{amsmath}
\usepackage{nicefrac}       
\usepackage{microtype}      
\usepackage{lipsum}
\usepackage{graphicx}
\usepackage{subcaption}
\usepackage{float}
\usepackage{tikz}
\usepackage{enumitem}
\usepackage{multirow}
\usepackage{bm}
\usepackage{algorithm}
\usepackage{algpseudocodex}

\usepackage{amsmath}
\usepackage{amsthm}

\usepackage{xfrac}

\usepackage{siunitx}

\usepackage{hyperref}

\usepackage[
    backend=biber,      
    style=numeric,      
    sorting=none,       
    giveninits=true,    
    sortcites=true,     
]{biblatex}

\theoremstyle{definition}
\newtheorem{definition}{Definition}

\usepackage{listings}

\lstdefinelanguage{myc++} 
{
  language=C++,
  captionpos=b,
  frame=single,
  keywordstyle=\bfseries\ttfamily,
  basicstyle=\scriptsize\ttfamily,
  commentstyle=\color{gray}\ttfamily,
  stringstyle=\rmfamily,
  numbers=left,
  numberstyle=\scriptsize,
  stepnumber=1,
  numbersep=8pt,
  breaklines=true,
  lineskip=1pt,
  frame=L,
  escapechar=\$,
  morekeywords = {constexpr},
}
\lstdefinelanguage{introspection} 
{
  captionpos=b,
  frame=single,
  keywordstyle=\bfseries\ttfamily,
  basicstyle=\scriptsize\ttfamily,
  commentstyle=\color{gray}\ttfamily,
  stringstyle=\rmfamily,
  numbers=left,
  numberstyle=\scriptsize,
  stepnumber=1,
  numbersep=8pt,
  breaklines=true,
  lineskip=1pt,
  frame=L,
  escapechar=\$
}

\usepackage{float}
\newfloat{listing}{tbp}{lop}  
\floatname{listing}{Listing}

\title{Efficient Wall-Modelled Large Eddy Simulation of Rotors using Homogenized Lattice Boltzmann Methods}

\author{
Adrian Kummerländer \\
Lattice Boltzmann Research Group \\
Institute of Applied and Numerical Mathematics \\
Karlsruhe Institute of Technology \\
\texttt{kummerlaender@kit.edu}
\And
Shota Ito \\
Lattice Boltzmann Research Group \\
Institute for Mechanical Process Engineering and Mechanics \\
Karlsruhe Institute of Technology
\And
Maximilian Schecher \\
Lattice Boltzmann Research Group \\
Institute of Applied and Numerical Mathematics \\
Karlsruhe Institute of Technology
\And
Davide Dapelo \\
Department of Civil and Environmental Engineering\\
University of Liverpool
\And
Stephan Simonis \\
Lattice Boltzmann Research Group \\
Institute of Applied and Numerical Mathematics \\
Karlsruhe Institute of Technology
\And
Mathias J. Krause \\
Lattice Boltzmann Research Group \\
Institute of Applied and Numerical Mathematics \\
Institute for Mechanical Process Engineering and Mechanics \\
Karlsruhe Institute of Technology
\And
Fedor Bukreev \\
Lattice Boltzmann Research Group \\
Institute for Mechanical Process Engineering and Mechanics \\
Karlsruhe Institute of Technology
}

\usepackage{xspace}
\usepackage{xcolor}
\usepackage{listings}

\lstdefinelanguage{myc++} 
{
  language=C++,
  captionpos=b,
  frame=single,
  keywordstyle=\bfseries\ttfamily,
  basicstyle=\scriptsize\ttfamily,
  commentstyle=\color{gray}\ttfamily,
  stringstyle=\rmfamily,
  numbers=left,
  numberstyle=\scriptsize,
  stepnumber=1,
  numbersep=8pt,
  breaklines=true,
  lineskip=1pt,
  frame=L,
  escapechar=\$
}
\newcommand*{\CodeComment}[1]{\hfill\makebox[7.0cm][l]{// #1}}

\definecolor{green2}{HTML}{00A36C}

\begin{document}

\maketitle

\begin{abstract}
Accurately capturing the dynamic forces acting on rotors as well as their wake effects presents a significant challenge for computational fluid dynamics (CFD) due to high Reynolds numbers and a large range of spatio-temporal scales. The present work proposes a novel blade-resolved wall-modeled large eddy simulation (WMLES) approach based on the lattice Boltzmann method (LBM).

A homogenized hybrid regularized recursive collision scheme targeting the filtered Brinkman--Navier--Stokes equations is combined with a novel wall-model. This is implemented in the context of a platform-transparent framework for fluid-structure interaction in the open source LBM framework OpenLB.

Convergence order and accuracy are validated against both experimental and numerical data for a model wind turbine, demonstrating excellent agreement for integral forces and wake velocity profiles. Computational efficiency and parallel scalability was investigated by roofline analysis and weak scaling studies for up to 384 rotors resolved by 41 billion lattice cells on the Karolina supercomputer.

The proposed framework enables efficient blade-resolved WMLES of entire wind farms and offers a new methodology for other complex wall-modeled fluid-structure interaction applications.
\end{abstract}

%

\section{Introduction}

Simulating fluid-rotor interaction is a fundamental challenge in aerospace, energy, and process engineering, governing the performance of systems from helicopter rotors and stirring tanks to wind turbines.
Analyzing these systems requires a choice among a spectrum of modeling fidelities~\cite{schaffarczyk2014}.
While low-order models like \emph{blade element momentum} (BEM), \emph{actuator disk} (AD) or line (AL) methods~\cite{hansen2006} can be sufficent for performance estimation, resolving the detailed, unsteady flow physics essential for load analysis, noise prediction, and design optimization demands high-fidelity blade-resolved \emph{computational fluid dynamics} (CFD)~\cite{oggiano2014,mittal2016,kirby2018,grinderslev2020,deoliveira2022}.

\emph{Direct numerical simulation} (DNS) of full-scale wind turbines remains infeasible due to the commonly very high Reynolds numbers (\(\sim 10^6 - 10^7\)) and large range of scales.
Instead, simulations commonly employ turbulence modeling via \emph{large eddy simulation} (LES).
In LES, only turbulent eddies above a threshold are resolved while smaller eddies are modeled, providing a compromise between accuracy and cost.
Additionally, in turbulent aerodynamics at high Reynolds numbers, plain LES approaches do not correctly capture the wall interactions in the boundary layer. As the features of the turbulence close to the wall are very different from other flow regions, dedicated handling of the near-wall region in a wall-modeled LES (WMLES) approach is required.
Such approaches approximate the near-wall region with empirically obtained wall functions.

The \emph{lattice Boltzmann method} (LBM) is a mesoscopic approach to the simulation of various transport phenomena in CFD \cite{kruger2017}.
Common target equations include the \emph{Navier-Stokes} (NSE), \emph{(reaction-)advection-diffusion} ((R)ADE) and radiative transport equations~\cite{simonis2023pde,mink2022,simonis23,bukreev2023}.
LB methods are particularly suited to massively parallel processing due to their algorithmic structure which can be split into a perfectly parallel \emph{collision step} and a neighborhood local \emph{streaming step}.

LBMs are established~\cite{korb2025} for both actuator line~\cite{rullaud2018,asmuth2020a,asmuth2022,schottenhamml2022,ribeiro2024,ribeiroBladeresolvedActuatorLine2025} and blade-resolved~\cite{asmuth2022,ribeiroBladeresolvedActuatorLine2025,xu2016,deiterding2016} LES of wind turbines and other rotors.
Considering blade-resolved LBM simulations as a subdomain of general-purpose \emph{fluid structure interaction} (FSI) problems, there is a rich variety approaches available to model the foundational moving boundary interactions.
A common way to model rotating geometries such as propellers or turbine blades is the coupling of sliding or overset meshes~\cite{yoo_hybrid_2021, yoo_compressible_2023, hedayat_parallel_2022, ribeiroBladeresolvedActuatorLine2025}.
More general approaches include \emph{immersed boundary methods} (IBM)~\cite{kruger2011,afra2018,wu2019,ye2020,fringand2024,rajamuni2024}, \emph{interpolated bounce back} (IBB)~\cite{bouzidi2001,obrecht2013,kaneda2014,kruger2017,haussmann2020,marson2021}, \emph{partially saturated} (PSM)~\cite{kruger2017,haussmann2020} and \emph{homogenized lattice Boltzmann methods} (HLBM)~\cite{krause_particle_2017,trunk_towards_2018,trunk2018,simonis23}.

The aim of the present work is to demonstrate that HLBMs are well suited to general FSI applications beyond particulate flows and that their locality and efficiency advantages translate to wall-modeled rotor flow simulations specifically.
To this end, the present study introduces a novel \emph{hybrid homogenized regularized recursive lattice Boltzmann fluid-structure interaction} (HHRRLBM-FSI) approach.
We detail its efficient implementation on GPU-accelerated high-performance computers (HPC) (Sections~\ref{sec:methodology} and \ref{sec:implementation}) and benchmark its accuracy (Section~\ref{sec:validation}) and performance (Section~\ref{sec:performance}) through application to the blade-resolved simulation of wind turbine aerodynamics -- a demanding test case for any FSI method.

\begin{figure}
  \includegraphics[width=\textwidth]{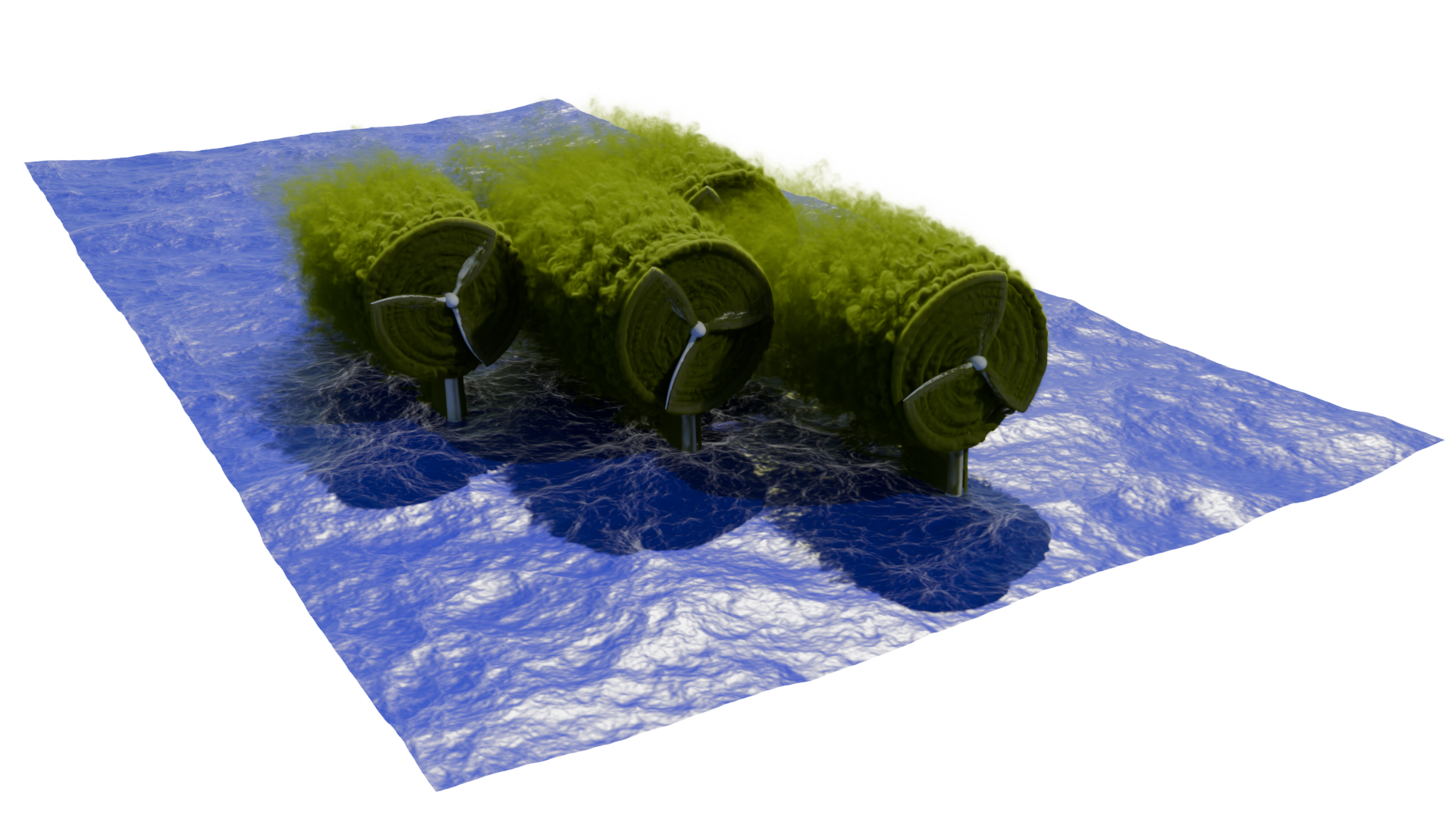}
  \caption{Illustrative volumetric rendering of the vorticity generated by a four-turbine offshore wind farm modeled as a two-way coupled blade-resolved LES in OpenLB.}
\end{figure}

\section{Methodology}\label{sec:methodology}


Modeling wind turbines is challenging for \emph{computational fluid dynamics} (CFD) due to the combination of large (landscape, the turbines themselves) and small length scales (boundary layer, blade surfaces).
In the present study we model this complex problem using the filtered Brinkman--Navier--Stokes equations (FBNSE), capturing both LES for subgrid-scale turbulence modeling and homogenized porous media for moving solid modeling in a single target equation.
This equation was solved efficiently~\cite{kummerlaender2023} in multi-GPU simulations using a \emph{homogenized lattice Boltzmann method} (HLBM) discretization~\cite{krause_particle_2017,simonis23} in OpenLB~\cite{krause2021,kummerlaender2025}.

\subsection{Filtered Brinkmann--Navier--Stokes Equations}

The macroscopic motion of fluids is commonly described using the Navier--Stokes equations (NSE).
Incompressible flows in heterogeneous domains consisting of both fully fluid regions and porous media can be described using the filtered Brinkman--Navier--Stokes equations (FBNSE)
\begin{align}
\begin{cases}
\bm{\nabla} \cdot \bar{\bm{u}}  =0, & \quad \text{in } \Omega \times I, \\
\frac{\partial \bar{\bm{u}}}{\partial t} + \bar{\bm{u}} \cdot \bm{\nabla} \bar{\bm{u}} = -\frac{\bm{\nabla} \bar{p}}{\rho} + \nu_{\mathrm{mo}} \bm{\nabla}^2 \bar{\bm{u}} + \frac{\nu_{\mathrm{mo}}}{K} \bar{\bm{u}} - \bm{\nabla} \cdot \mathbf{T}_{\mathrm{sgs}}, & \quad \text{in } \Omega \times I, \label{eq:fbnse}
\end{cases}
\end{align}
for filtered pressure \(\bar{p}\), velocity \(\bar{\bm{u}}\) density \(\rho\) 
on spatial domain \(\Omega \subseteq \mathbb{R}^3\) and time \(I\subseteq \mathbb{R}_{>0}\).
The molecular kinematic viscosity is defined as \(\nu_{\mathrm{mo}}\) and the permeability coefficient $K>0$ of the porous medium is given by the Forchheimer equation
\begin{equation}
K = \frac{\mu_F Q}{A \left(\frac{\triangle P}{L} - \frac{\rho}{K_{\beta}} \frac{Q^2}{A^2} \right)},
\end{equation}
with dynamic viscosity $\mu$, volume flow rate $Q$, characteristic length $L$, projected area $A$, pressure difference $\triangle P$ and nonlinear permeability coefficient $K_\beta$.
The term $\bm{\nabla} \cdot \mathbf{T}_{\mathrm{sgs}}$ models the subgrid-scale turbulence using the Smagorinsky LES approach  
\begin{align}
    \mathbf{T}_{\mathrm{sgs}} & = 2 \nu_\mathrm{turb} \bar{\mathbf{S}}, \label{eq:sgsStress}\\
    \nu_\mathrm{turb} & = \left(C_{\mathrm{S}} \triangle x \right)^2 \left|\bar{\mathbf{S}}\right|, \label{eq:turbVisc}
\end{align}  
where \(C_{\mathrm{S}}>0\) is the Smagorinsky constant, \(\triangle x\) is the filter width, and $\bar{\mathbf{S}}$ is the filtered strain rate tensor:  
\begin{equation}
    \bar{S}_{\alpha\beta} = \frac{1}{2} \left( \frac{\partial \bar{u}_{\alpha}}{\partial x_\beta} + \frac{\partial \bar{u}_{\beta}}{\partial x_\alpha} \right).
\end{equation} 

\subsection{Homogenized Lattice Boltzmann Method}\label{sec:hhrrlbm}

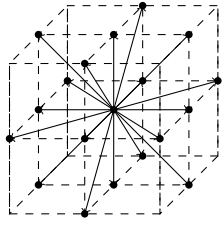
\begin{figure}
\centering
\begin{tikzpicture}[scale=1, every node/.style={circle, fill=black, inner sep=1pt}]
            \draw[dashed] (-1, -1, -1) -- (-1, -1, 1) -- (-1, 1, 1) -- (-1, 1, -1) -- cycle; 
            \draw[dashed] (1, -1, -1) -- (1, -1, 1) -- (1, 1, 1) -- (1, 1, -1) -- cycle;   
            \draw[dashed] (-1, -1, -1) -- (1, -1, -1) -- (1, 1, -1) -- (-1, 1, -1) -- cycle; 
            \draw[dashed] (-1, -1, 1) -- (1, -1, 1) -- (1, 1, 1) -- (-1, 1, 1) -- cycle;   
            \draw[dashed] (0, -1, -1) -- (0, -1, 1) -- (0, 1, 1) -- (0, 1, -1) -- cycle;
            \draw[dashed] (-1, 0, -1) -- (-1, 0, 1) -- (1, 0, 1) -- (1, 0, -1) -- cycle;
            \draw[dashed] (-1, -1, 0) -- (-1, 1, 0) -- (1, 1, 0) -- (1, -1, 0) -- cycle;

            \foreach \x/\y/\z in {0/0/0, 1/0/0, -1/0/0, 0/1/0, 0/-1/0, 0/0/1, 0/0/-1,
                                  1/1/0, -1/-1/0, 1/-1/0, -1/1/0, 1/0/1, -1/0/-1,
                                  0/1/1, 0/-1/-1, 1/0/-1, -1/0/1, 0/1/-1, 0/-1/1}
                \node at (\x, \y, \z) {};

            \draw[->] (0, 0, 0) -- (1, 0, 0);
            \draw[->] (0, 0, 0) -- (-1, 0, 0);
            \draw[->] (0, 0, 0) -- (0, 1, 0);
            \draw[->] (0, 0, 0) -- (0, -1, 0);
            \draw[->] (0, 0, 0) -- (0, 0, 1);
            \draw[->] (0, 0, 0) -- (0, 0, -1);
            \foreach \x/\y/\z in {1/1/0, -1/-1/0, 1/-1/0, -1/1/0, 1/0/1, -1/0/-1,
                                  0/1/1, 0/-1/-1, 1/0/-1, -1/0/1, 0/1/-1, 0/-1/1}
                \draw[->] (0, 0, 0) -- (\x, \y, \z);
        \end{tikzpicture}
\caption{Schematic of the discrete velocity set \(D3Q19\).}
\label{fig:lattice}
\end{figure}

The HLBM is used to discretize the FBNSE~\eqref{eq:fbnse} on a regular space-time grid with the \(D3Q19\) velocity stencil (cf. Figure~\ref{fig:lattice}).
Specifically, we utilize a \emph{homogenized hybrid regularized recursive lattice Boltzmann method with Smagorinsky LES model} (HHRRLBM-LES) that extends the classic HLBM~\cite{krause_particle_2017} with a hybrid third-order recursive regularized collision model~\cite{coreixas2017,JACOB18}.

The filtered and homogenized LB equation is given by  
\begin{equation}\label{eq:hlbm}
    f_{i} (\bm{x}+\xi_i \triangle t, t+\triangle t) 
    = 
    f_{i}^{\mathrm{eq}} (\bm{x}, t) + \left( 1 - \frac{1}{\tau_{\mathrm{eff}}(\bm{x},t)} \right)\tilde{f}_{i}^{(1)}(\bm{x}, t), \quad \text{in } \Omega_{\triangle x} \times I_{\triangle t}, 
\end{equation}
for distribution functions \(f_i\) along \(q\) discrete velocities \(\xi_i\) on a regular lattice \(\Omega_{\triangle x} \subset \Omega \subseteq \mathbb{R}^3\) with cell size \(\triangle x\) at discrete times \(I_{\triangle t} \subset I \subseteq \mathbb{R}_{\geq 0}\) separated by step size \(\triangle t\).
Here, the non-equilibrium distribution \(\tilde{f}_i^{(1)}\) is computed as a linear combination
\begin{align}
    \tilde{f}_{i}^{(1)}(\bm{x},t) = \sigma f_{i}^{(1)} (\bm{x},t)  - (1-\sigma) f_{i}^{(1,FD)} \text{ for } \sigma \in [0,1].
\end{align}
That is, the distribution is \emph{hybridized} between reconstructions using the rate of strain tensor obtained either from local macroscopic moments or from a non-local finite difference (FD) approximation.

For the local part, the non-equilibrium distribution function $f_{i}^{(1)}$ is expanded in terms of Hermite polynomials \(\mathbf{H}_i^{(n)}\) of the discrete velocity \(\xi_i\) as
\begin{equation}
  f_{i}^{(1)}(\bm{x},t) = \omega_i \sum_{n=0}^{N=3} \frac{1}{c_{\mathrm{s}}^{2n} n!} \mathbf{H}_i^{(n)} : \mathbf{a}_1^{(n)}(\bm{x},t) ,
\end{equation}
where \(\omega_i\) are the lattice weights. The Hermite expansion coefficients are defined as
\begin{equation}
    \mathbf{a}_1^{(n)}(\bm{x},t)=\sum_{i=0}^{q-1}\mathbf{H}_i^{(n)}f_i^{(1)}(\bm{x},t).
\end{equation}
For the non-local part, the FD non-equilibrium distribution function is defined as
\begin{equation}
f_{i}^{(1,FD)} := \frac{\rho \tau}{c_{\mathrm{s}}^{2}} \mathbf{H}_{i}^{(2)} : \mathbf{S}^{\mathrm{FD}} (\bm{x},t).
\end{equation}

The equilibrium distribution function is defined as
\begin{equation}
  f_i^{\mathrm{eq}}(\bm{x},t) = \omega_i \left(\rho + \frac{\xi_i \cdot \rho\, \mathbf{\widehat{u}}}{c_{\mathrm{s}}^2}
    + \frac{\mathbf{H}_i^{(2)} : \widehat{\mathbf{a}}_0^{(2)}}{2 c_{\mathrm{s}}^4}
    + \frac{\mathbf{H}_i^{(3)} : \widehat{\mathbf{a}}_0^{(3)}}{2 c_{\mathrm{s}}^6}
  \right)
\end{equation}
using Hermite coefficients \(\widehat{\mathbf{a}}_{0}^{(0)} = \rho(\bm{x},t)\) and \(\widehat{\mathbf{a}}_{0}^{(n)} = \mathbf{a}_{0}^{(n-1)} \widehat{\bm{u}}(\bm{x},t)\).

In the general case~\cite{krause_particle_2017}, we define the homogenized velocity \(\widehat{\bm{u}}\) as a convex combination of the fluid velocity moment \(\bm{u}\) and the solid velocity \(\bm{u}^{\mathrm{B}}\), given by  
\begin{equation}\label{eq:convexVelocity}
    \widehat{\bm{u}}(\bm{x},t) = (1 - d(\bm{x},t)) \bm{u}(\bm{x},t) + d(\bm{x},t) \bm{u}^{\mathrm{B}}(\bm{x},t), 
\end{equation}
where \(d\) is the so-called lattice porosity
\begin{equation}
    d(\bm{x},t) = 1 - \frac{\triangle x^2 \nu \tau_{\mathrm{mo}}}{K(\bm{x},t)}. 
    \label{eq:lattice_porosity}
\end{equation}
and \(\tau_{\mathrm{mo}}\) is the molecular relaxation time.
To ensure a non-slip boundary condition on the moving solid surface, \(\bm{u}^\mathrm{B}\) must be set to the velocity of the solid surface. Otherwise, momentum is not transferred from the solid.
Section~\ref{sec:wm} describes how this is combined with dynamically-prescribed correction velocites in the wall model.

Finally, the subgrid scale turbulence is accounted for by locally computing the effective relaxation time \(\tau_\mathrm{eff}(\bm{x},t)\) using the Smagorinsky BGK model
\begin{equation}\label{eq:tauEff}
    \tau_\mathrm{eff}(\bm{x},t) = \frac{\nu_\mathrm{eff}(\bm{x},t)}{c_{\mathrm{s}}^2} \frac{\triangle t}{\triangle x^2} + \frac{1}{2}.
\end{equation}

Connecting the HHRRLBM~\eqref{eq:hlbm} to the FBNSE~\eqref{eq:fbnse} target equation, we expect a second-order approximation in space for the fluid velocity moment \cite{simonis23,simonis2023pde}.

\subsubsection{Turbulent Wall Model}\label{sec:wm}

\begin{figure}
  \begin{center}
  \includegraphics[width=0.5\textwidth,trim={0 0.5cm 0 0}]{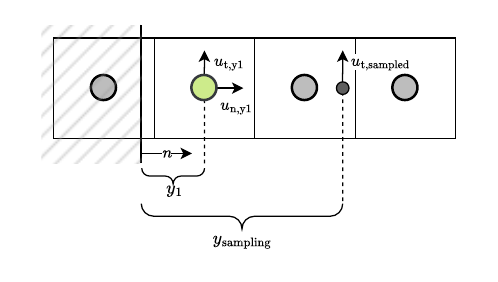}
  \end{center}
  \caption{Schematic of the wall-modeled region. Wall-modeled cell marked in green.}
  \label{fig:wallfunction}
\end{figure}

It is computationally very expensive to a degree that renders it \emph{infeasible} to fully resolve the steep velocity gradients characterizing the boundary layer flow around full-scale wind turbines' blade surfaces.
An established solution to this problem is the use of explicit wall modeling s.t. the velocity profile in the boundary layer are approximated using \emph{universal turbulent velocity profiles}~\cite{karman1930,wilcox1998} instead of being resolved by the mesh.
This holds both for CFD in general and LBM in particular~\cite{malaspinas2014}.
Given a kinematic fluid viscosity \(\nu\), a wall distance \(y\) and wall shear stress \(\tau_\text{w}\) the near-wall velocity \(u\) is de-dimensionalized as
\begin{equation}\label{eq:u+}
    u^+ = u \sqrt{\frac{\rho}{\tau_w}} = \frac{u}{u_{\tau_w}},
\end{equation}
and the distance to the wall \(y\) as
\begin{equation}\label{eq:y+}
    y^+ = \frac{y}{\nu}\sqrt{\frac{\tau_w}{\rho}} = y \frac{u_{\tau_w}}{\nu}.
\end{equation}
Empirical equations relating these two quantities in the near-wall region are referred to as \emph{wall functions}.
For the present case, we use the established truncated form of the Spalding wall function~\cite{spalding1961}
\begin{equation}\label{eq:spalding}
    y^+ = u^+ + \frac{1}{E}\left( \exp{\left(\kappa u^+\right)} - 1 - \kappa u^+ - \frac{(\kappa u^+)^2}{2} - \frac{(\kappa u^+)^3}{6} \right),
\end{equation}
with empirical parameter \(E \approx 9.8\) and von K{\'a}rm{\'a}n constant \(\kappa \approx 0.41\).
The resulting implicit equation
\begin{equation}\label{eq:implicitWF}
    0 = f\left(u_{\tau_w}\right) = \frac{u}{u_{\tau_w}} + \frac{1}{E}\left( \exp{\left(\kappa \frac{u}{u_{\tau_w}}\right)} - 1 - \kappa \frac{u}{u_{\tau_w}} - \frac{1}{2}\left(\kappa \frac{u}{u_{\tau_w}}\right)^2 - \frac{1}{6}\left(\kappa \frac{u}{u_{\tau_w}}\right)^3 \right) - y \frac{u_{\tau_w}}{\nu}
\end{equation}
is solved for the friction velocity \(u_{\tau_w}\) using a Newton-Raphson iteration method
\begin{equation}
  u_{\tau_w}^{j+1} := u_{\tau_w}^j - \frac{f(u_{\tau_w}^j)}{f'(u_{\tau_w}^j)}
\end{equation}
given an initial guess
\begin{equation}\label{eq:guess}
  u_{\tau_w}^{j=0} := \begin{cases}u_{\tau_w}^\text{prev}&\text{wall function active during previous timestep}\\\sqrt{\nu\frac{|u_\text{t,sampled}|}{\|y_\text{sampling}\|}}&\text{otherwise}\end{cases}.
\end{equation}
For the case where no value from the previous timestep is available, the initial guess in Equation~\ref{eq:guess} is computed using interpolated velocity at configurable position \(y_\text{sampling}\).
Figure~\ref{fig:wallfunction} illustrates the involved cells at the solid boundary.
In the present case this iteration is executed until either \(|f(u_{\tau_w}^{j+1})-f(u_{\tau_w}^j)| < \num{1e-4}\) is satisfied or for a maximum of 10 iterations.

The resulting modeled tangential friction velocity \(u_{\tau_w}\) is incorporated into the HHRRLBM by adding it to the velocity moment of the porous medium.
Finally, shear stresses are corrected by setting the hybridization factor \(\sigma=0\) for all wall-modeled cells to increase stability at the moving boundary.

Section~\ref{sec:implementation} details how this wall model is implemented efficiently in OpenLB.

\subsubsection{Fluid-Structure Interaction}

In the present work, we use the FBNSE (cf. Equation~\ref{eq:fbnse}) realized via (HH)RRLBM for the whole simulation domain except the outer boundary conditions (cf. reference setup in Figure~\ref{fig:domain}).
This allows incorporation of both moving and static resolved solid geometries without remeshing.

The lattice porosity defined in Equation~\ref{eq:lattice_porosity} is stricly not grid-independent, being computed from spatial discretization, fluid viscosity, molecular relaxation time and permeability. As such, rigorously connecting the lattice porosity to a macroscopic solid geometry is challenging~\cite{ito2025}.
For the established usage of HLBM in resolved particulate flows~\cite{krause_particle_2017,trunk_towards_2018,trunk2018} this was approached by considering the lattice porosity as a function of the signed distance to the solid surface.
\begin{definition}[Lattice Porosity]
Let \(\phi(\mathbf{x},t)\) be the signed distance to the solid surface, \(\epsilon_h=\epsilon \triangle x\) the width of the smooth transition region coupled to spatial discretization factor \(\triangle x\) and \(s : \mathbb{R} \to [0,1]\) a transition function. Then
\begin{equation}
  d(\mathbf{x},t) := \begin{cases}
    0 & \text{if } \phi(\mathbf{x},t) \le -\frac{\epsilon_h}{2} \\
    s(\phi(\mathbf{x},t)) & \text{if } \phi(\mathbf{x},t) \in (-\frac{\epsilon_h}{2}, \frac{\epsilon_h}{2}) \\
    1 & \text{if } \phi(\mathbf{x},t) \ge \frac{\epsilon_h}{2}
\end{cases}
\end{equation}
is the lattice porosity. A common choice for the transition function \(s\) is
\begin{equation}
  s(\phi) = \frac{\phi}{\epsilon_h} + \frac{1}{2},
\end{equation}
s.t. the \emph{real} solid wall is the level set of \(d(\mathbf{x},t)=\sfrac{1}{2}\) (cf. \cite[Figure~2]{krause_particle_2017}).
\end{definition}
The error introduced by the disconnect between physical permeability definition and lattice permeability is acceptable as we achieve a sharp interface for \(\triangle x \to 0\) due to the coupling of transition width and spatial resolution.

While application of HHRRLBM together with the wall model detailed in Section~\ref{sec:wm} is sufficient to represent boundary motion (the \emph{structure response}), full FSI coupling also requires exchange of information from the fluid to the solid. For the present work we require the (integral) forces imposed by the fluid on the rotor geometry.
An established approach is to compute the momentum exchange~\cite{wen2014} in the local neighborhood of the boundary.
\begin{equation}
\mathbf{F}(t) = \sum_{\{\mathbf{x} | d(\mathbf{x},t) < 1\}} \sum_{i=1}^{q-1} \left[ (\mathbf{c}_i - \mathbf{u}_s)f_i(\mathbf{x} + \mathbf{c}_i, t) + (\mathbf{c}_i + \mathbf{u}_s)f_i(\mathbf{x}, t) \right].
\end{equation}

\section{Implementation}\label{sec:implementation}

The HHRRLBM~\eqref{eq:hlbm} collision step detailed in Section~\ref{sec:hhrrlbm} and used in the wall-modeled regions of the simulation domain is implemented using OpenLB's dynamics tuple system, a kind of \emph{domain specific language} (DSL) for local LB cell models~\cite{kummerlaender2023,kummerlaender25}.
Listing~\ref{lst:hrrlbm} shows how the non-wall-modeled bulk is constructed as a tuple of moment system, equilibrium and collision operator.
\begin{listing}[b]
\begin{lstlisting}[language=myc++]
using HHRRLBM = dynamics::Tuple<
  T, descriptors::D3Q19<>, $\Comment{Value type and lattice stencil}$
  typename momenta::Tuple< $\Comment{Macroscopic moment system}$
    momenta::BulkDensity,
    momenta::MovingPorousMomentumCombination<momenta::BulkMomentum>,$\Comment{HLBM via moment system}$
    momenta::BulkStress,
    momenta::DefineToNEq
  >,
  equilibria::ThirdOrder, $\Comment{Equilibrium distribution}$
  collision::ParameterFromCell<$\Comment{Modified collision operator}$
    collision::HYBRID, 
    collision::SmagorinskyEffectiveOmega<collision::HRR>>
>;
\end{lstlisting}
\caption{Dynamics tuple formulation of the HHRRLBM-LES scheme used for wall-modeled cells}
\label{lst:hrrlbm}
\end{listing}
To reduce the arithmetic and bandwidth requirements a simpler RRLBM-LES model is applied for all parts of the domain that are guaranteed to not intersect the moving wall-modeled structures.
Listing~\ref{lst:rrlbm} shows how this model is described as a dynamics tuple.
\begin{listing}[b]
\begin{lstlisting}[language=myc++]
using RRLBM = dynamics::Tuple<
  T, descriptors::D3Q19<>, $\Comment{Value type and lattice stencil}$
  typename momenta::Tuple< $\Comment{Macroscopic moment system}$
    momenta::BulkDensity,
    momenta::BulkMomentum,
    momenta::BulkStress,
    momenta::DefineToNEq
  >,
  equilibria::ThirdOrder, $\Comment{Equilibrium distribution}$
  collision::ParameterFromCell<$\Comment{Modified collision operator}$
    collision::LES::SMAGORINSKY,
    collision::SmagorinskyEffectiveOmega<collision::ThirdOrderRLB>>
>;
\end{lstlisting}
\caption{Dynamics tuple formulation of the RRLBM-LES scheme used for bulk cells}
\label{lst:rrlbm}
\end{listing}
We refer to Section~\ref{sec:performance} for an analysis of the resulting performance of these local models.

\subsection{Representation of Moving Geometries}\label{sec:georep}

Similarly to the established~\cite{krause_particle_2017,trunk_towards_2018,trunk2018} modeling of arbitrarily shaped resolved particles using HLBM, we voxelize the moving geometry elements in a separate inertial reference grid.
This separate grid is in fact realized using the exact same lattice data structure that is used for the LBM, laying the foundation for utilizing recent LBM-native structure solvers~\cite{boolakee2023_1,boolakee2023_2,mueller2025} for modeling elastic structure deformation.
For the present first application of our wall-modeled approach, only rotation is considered and the reference grid is resolved at double the spatial resolution of the fluid lattice.

In any case, reference grids are embedded into HLBM lattice porosities for given locations and orientations at every discrete timestep.
Each per-cell porosity value is associated with a integer \emph{tag}, connecting it to the specific \emph{FSI element} it belongs to.
This way, obtaining the fluid response (e.g. integral forces) reduces to per-tag summation for which efficient parallel algorithms (reductions) are available~\cite{brent1979,bell2012,hwu2023}.

\begin{figure}
  \includegraphics[width=\linewidth,trim={0 7cm 0 0}]{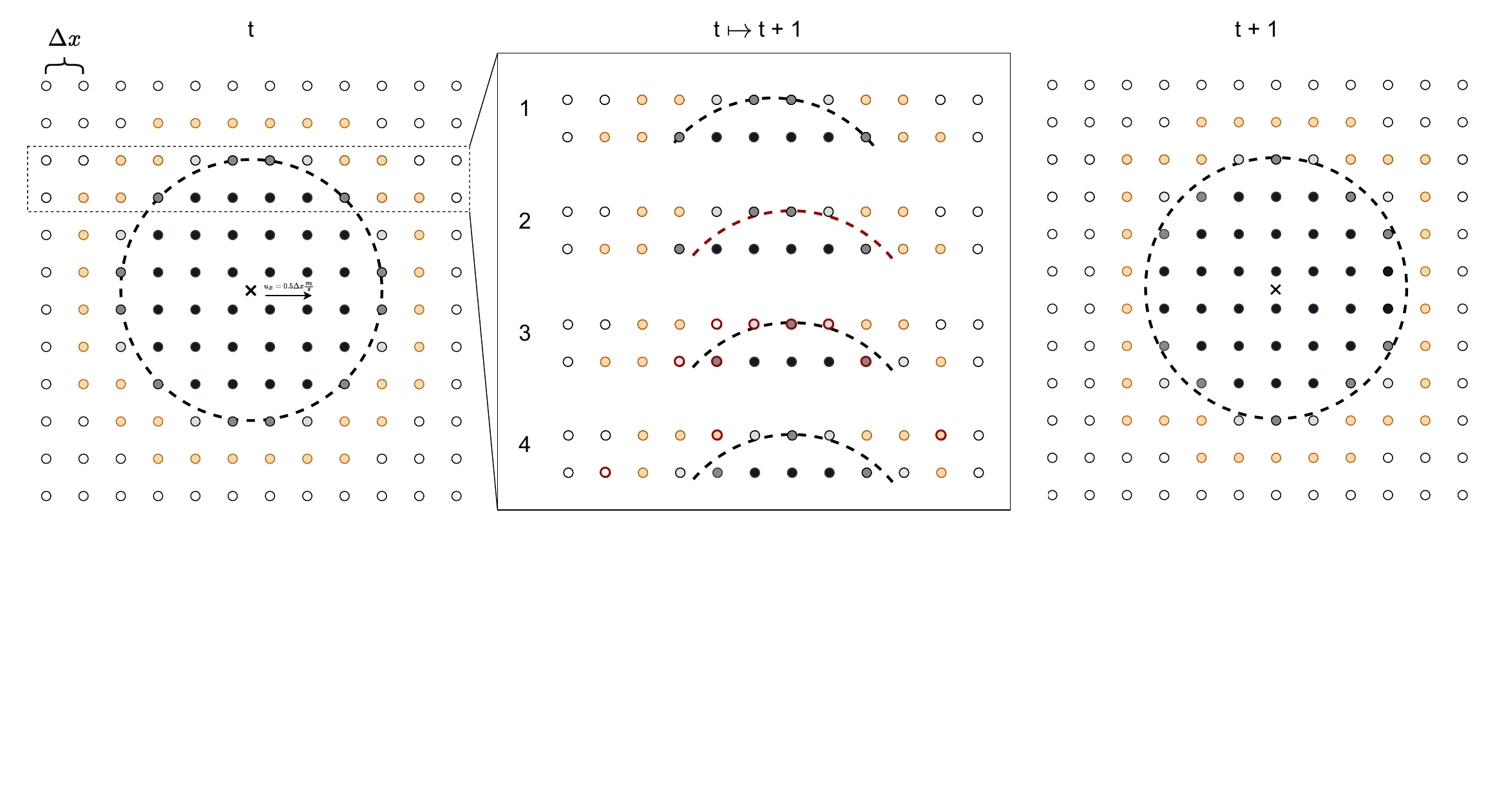}
  Dashed black circle marks the level set \(d(\mathbf{x},t)=\sfrac{1}{2}\).
  Black cells are fully in the solid.
  Shaded gray cells are at the solid frontier, this is where the momentum exchange takes place.
  Orange cells mark the fully-fluid growth layer.
  White cells are fully-fluid and separated by at least one cell from the FSI region.
  \begin{enumerate}
    \item Original state before structure update, tagged porosity field and growth layer aligned
    \item Update structure
    \item For each tagged cell re-compute the porosity embedding
    \item Reset tag for each full-fluid tagged cell without solid neighbor
  \end{enumerate}
  \caption{Efficient local update of porosities from signed distance geometry}
  \label{fig:fsialgo}
\end{figure}

Updating the porosities due to a element's displacement can be implemented efficiently without re-checking all cells due to boundary movement necessarily being bounded by the lattice speed of sound.
That is, between each discrete time step any surface may only displace less than the distance between adjacent cells.
Even more restricted, we can reasonably expect all solid movements to be bounded by the characteristic lattice velocity.
As such, we only need to check whether a element has moved into a single-cell-width \emph{growth layer} (cf. Figure~\ref{fig:fsialgo}) and, if so, ensure that this layer moves with the element.

\begin{figure}
    \begin{subfigure}{0.5\textwidth}
        \includegraphics[width=\linewidth]{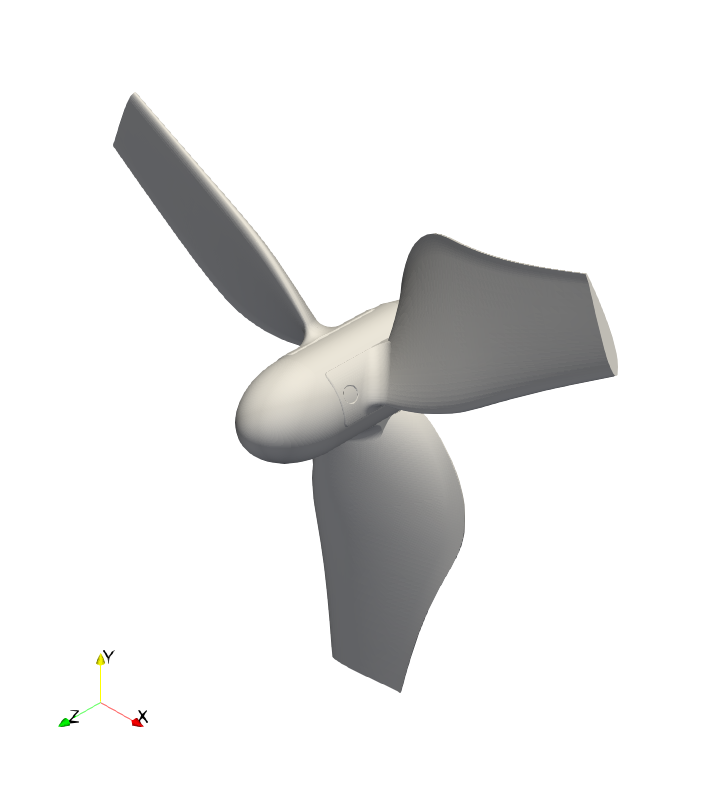}
        \caption{Cleaned rotor geometry as STL}
        \label{fig:rotorstl}
    \end{subfigure}
    \begin{subfigure}{0.5\textwidth}
        \includegraphics[width=\linewidth]{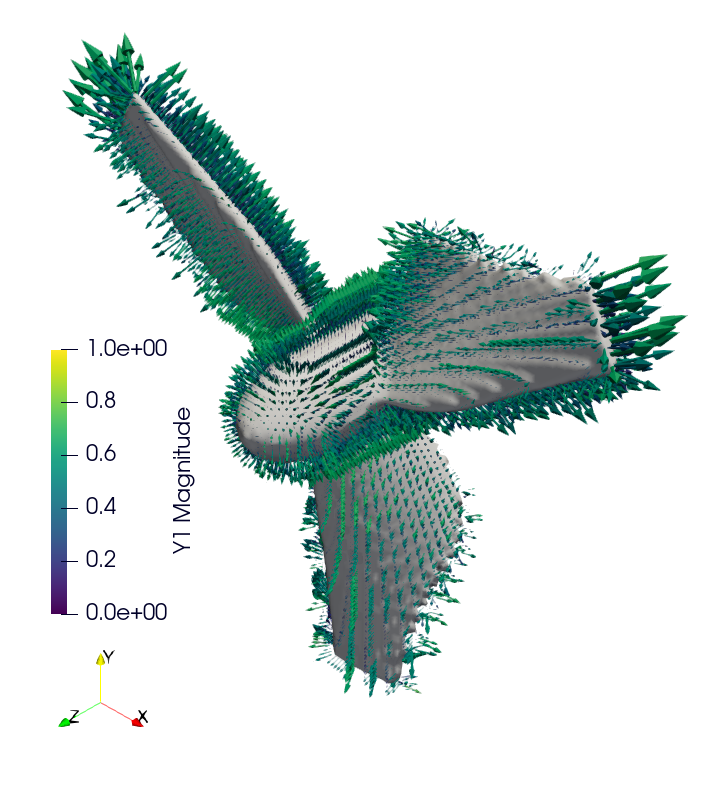}
        \caption{Level set of \(d(\mathbf{x})=\sfrac{1}{2}\) and \(\mathbf{y}_1\) surface normal vectors}
        \label{fig:y1stl}
    \end{subfigure}
    \caption{Representations of rotor geometry for wall-modeling}
\end{figure}

\subsection{Moving Wall Model}

Due to its non-local nature, the wall model detailed in Section~\ref{sec:wm} is implemented as OpenLB \emph{post processors} that are applied to the entire wall-modeled FSI region but only activate for cells at the porosity frontier by inspecting \(d(\mathbf{x},t)\). Analogously to local dynamics, these non-local operators are implemented against the concept of a cell and as such directly benefit from OpenLB's platform-transparent execution~\cite{kummerlaender2023,kummerlaender25} and are amenable to automatic code optimization.

Listing~\ref{lst:hhrrlbmfsi} summarizes the LB algorithm for the wall model's perspective. Compared to a non-wall-modeled case the main difference is the addition of a \texttt{WallModelO} post processor implementing both the Newton-Raphson iteration to approximate the friction velocity and the finite difference approximation of the shear stress (cf. Equation~\ref{eq:spalding}) given locally stored \(\mathbf{y}_1\) normals and the sampling distance parameter.
The \(\mathbf{y}_1\) normal vector (scaled by the normalized \(y_1\) wall distance) is updated for boundary cells during the lattice porosity update in \texttt{UpdatePorosityO}. For the present rotor case the resulting vectors at the surface are illustrated in Figure~\ref{fig:y1stl}.

\begin{listing}
\begin{algorithmic}[1]
\Statex \Call{InitializePorosityO}{} \CodeComment{One-time global porosity and Y1 field setup}
\Loop
\State \Call{UpdatePorosityO}{} \CodeComment{Update tagged porosities and Y1 in growth layer}
\State Collide and Stream \CodeComment{Usual (HH)RRLBM timestep}
\State \Call{WallModelO}{} \CodeComment{Compute wall modelled velocites and strain rate at Y1}
\State \Call{CollectPorousBoundaryForceO}{} \CodeComment{Compute per-cell boundary force at element frontier}
\State \Call{IntegratePorousBoundaryForceO}{} \CodeComment{Reduce per-element force}
\Statex \texttt{sort\_by\_key}, \texttt{reduce\_by\_key} \CodeComment{Most expensive part of FSI}
\State Evolve structure \CodeComment{Independent of LBM, various approaches e.g. DEM, FEM, Solid-LBM}
\EndLoop
\end{algorithmic}
\caption{HHRRLBM-FSI}\label{lst:hhrrlbmfsi}
\end{listing}

\section{Validation}\label{sec:validation}

To validate the present HHRRLBM-FSI approach for applications to wind power plants, we reproduce the experimental and blade-resolved simulation results for a three-bladed rotor by Ribeiro et al.~\cite{ribeiroBladeresolvedActuatorLine2025}.
We chose this study in particular because it also uses LBM, its strong experimental foundation with detailed validation data, and especially due to the open availability of the concrete rotor geometry which sets it apart from similar studies.

\begin{figure}
  \begin{center}
  \includegraphics[width=0.5\textwidth]{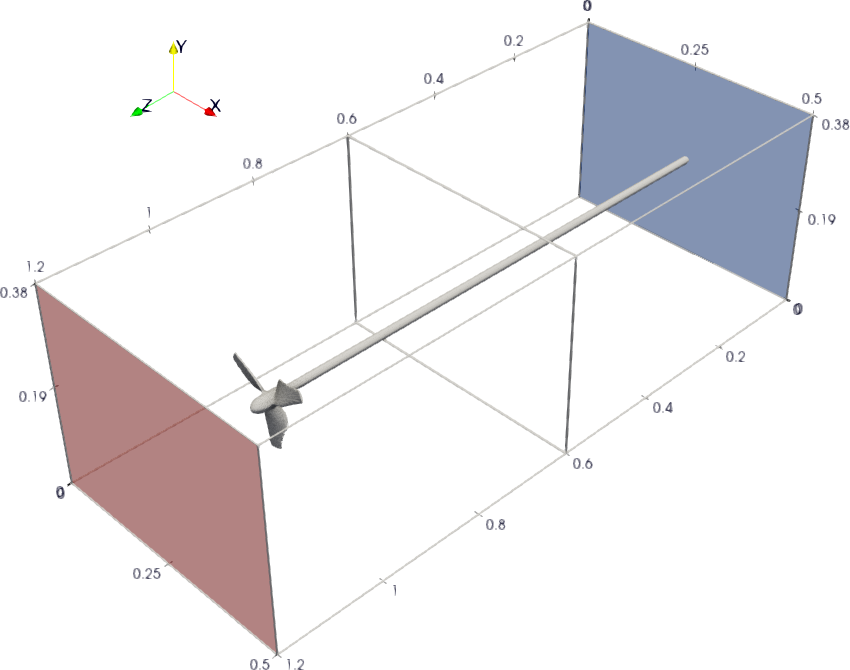}
  \end{center}
  \caption{Isometric sketch of the simulation domain. Inlet- and outlet planes are marked in red and blue respectively.}
  \label{fig:domain}
\end{figure}

%

The simulation domain depicted in Figure~\ref{fig:domain} is \qty{1.2}{\meter} long along the axial direction (z-axis) and spans \qty{0.5}{\meter} resp. \qty{0.38}{\meter}. The rotor (cf. Figure~\ref{fig:rotorstl}) with radius \(r := \qty{0.09}{\meter}\) and tip chord width \(c_\text{tip} := \qty{0.023}{\meter}\) is placed at the x-y center and \qty{0.96}{\meter} from the outlet towards the inflow. It is connected to a shaft with radius \(r_\text{shaft} := \qty{0.00725}{\meter}\) that extends downstream from the rotor to the outlet.
Both the rotor and the shaft always rotate as a single unit with the same terminal frequency \(f := \qty{3}{\hertz}\).
The free-stream axial velocity is set to \(U_\infty := \qty{0.56}{\meter\per\second}\).
Following Definition~\ref{def:reynolds}, the characteristic Reynolds number of the present system is \(\text{Re} := \num{38882}\) for kinematic viscosity \(\nu := \qty{1.0035e-6}{\meter\squared\per\second}\).

\begin{definition}[Tip chord based Reynolds number]\label{def:reynolds}
Let \(f\) [\si{\hertz}] be the rotational frequency of a rotor with radius \(r\) [\si{\meter}] and tip chord length \(c_\text{tip}\) [\si{\meter}]. Then
\begin{equation}
u_\text{tip} = 2\pi r f
\end{equation}
is the local tip velocity.
Given the kinematic viscosity \(\nu\) [\si{\meter\squared\per\second}] of a fluid,
\begin{equation}
\text{Re} := \frac{u_\text{tip} c_\text{tip}}{\nu}
\end{equation}
is the dimensionless tip chord based Reynolds number.
\end{definition}

\begin{figure}
  \centering
  \includegraphics[width=\textwidth]{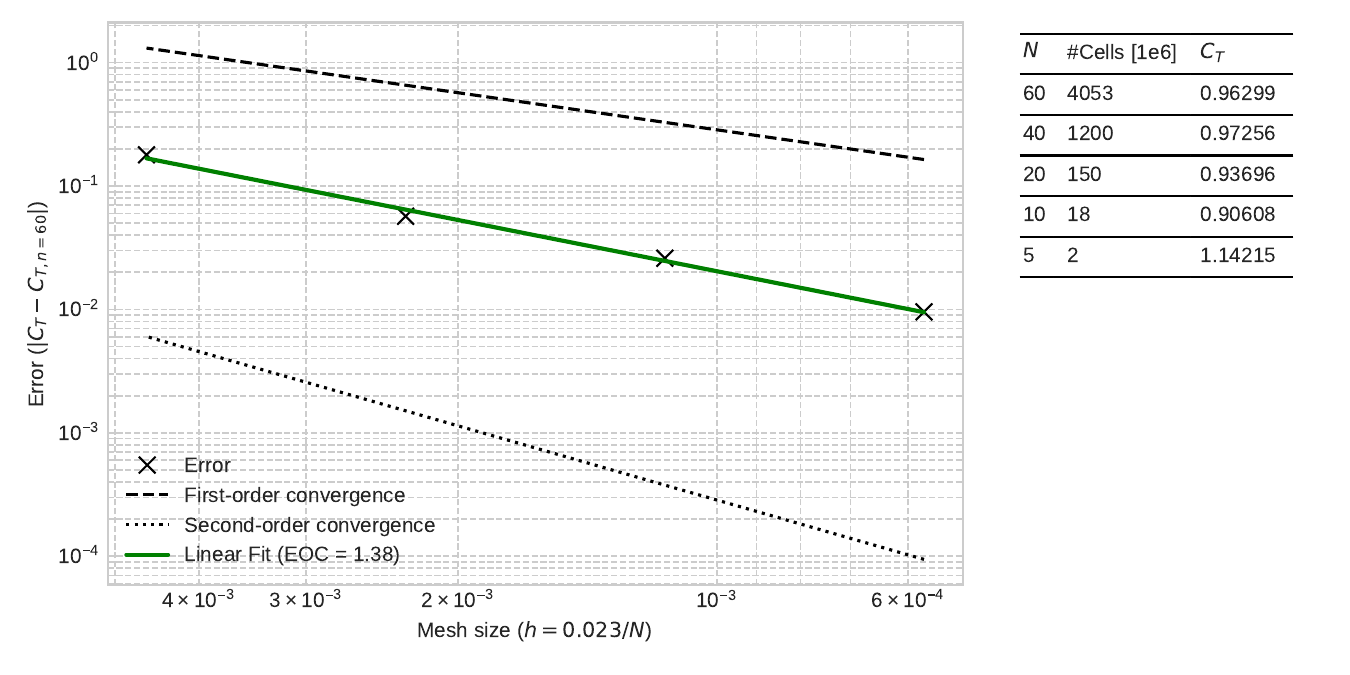}
  \caption{Grid Convergence Analysis of Thrust Coefficients}
  \label{fig:eoc}
\end{figure}

Different from the reference setup~\cite{ribeiroBladeresolvedActuatorLine2025}, our present simulation approach doesn't employ local grid refinement but relies on a uniform mesh with wall-modeled moving boundaries (cf. Section~\ref{sec:methodology}).
The uniform grid is parameterized by the spatial discretization factor \(\triangle x = \frac{c_\text{tip}}{N}\) and the time discretization \(\triangle t\) is computed to yield a characteristic lattice velocity \(u_\text{lattice} = 0.05\).
Relating the stated resolution of \(\Delta x_\text{wake} = r/225\) for the wake refined region to our present uniform mesh, this is equivalent to a tip chord resolution \(N=\num{57.5}\) which is similar to the finest simulated case of \(N=\num{60}\) (cf. Figure~\ref{fig:eoc}). The remaining difference between both approaches is that we employ a wall-modeled moving boundary approach while the reference resolves the boundary layer in a rotating reference frame using a overset grid approach.
The finest \(N=\num{60}\) case achieves a maximum \(y^+\approx\num{31}\) which is within the valid region of the employed Spalding wall function~\cite{spalding1961}.

\begin{definition}[Thrust Coefficient]\label{def:thrust}
Let \(\text{T}\) be the integral of all axial forces acting on the rotor, \(\rho, U_\infty\) and \(r \in \mathbb{R}^+\) the fluid density, characteristic free stream velocity resp. rotor radius. Then
\begin{equation}
\text{C}_\text{T} := \frac{\text{T}}{0.5 \rho \pi r^2 U_\infty^2}
\end{equation}
defines the non-dimensional \emph{thrust coefficient}.
\end{definition}

Figure~\ref{fig:eoc} shows the results of a grid convergence study conducted on the Karolina supercomputer at IT4I.
Five different grid resolutions \(N = \{5, 10, 20, 40, 60\}\) corresponding to grids between 2 million and 4 billion cells were analyzed.
The absolute error for the time-averaged thrust coefficient \(\text{C}_\text{T}\) (cf. Definition~\ref{def:thrust}) was calculated for each resolution with respect to the finest thrust coefficient of \(\text{C}_\text{T} \simeq 0.963\).
Figure~\ref{fig:eoc} displays the log-log plot of this error as a function of the characteristic mesh size, \(h=0.023/N\).
The data exhibits a clear linear trend, indicating that the solution is within the asymptotic range of convergence.
A linear regression of the errors yields an \emph{experimental order of convergence} (EOC) of 1.38.

\begin{definition}[Grid Convergence Index]\label{def:gci}
Let \(f_1, f_2 \in \mathbb{R}\) be integral quantities of interest from a numerical simulation, obtained on a grids with characteristic mesh sizes \(h_1, h_2 \in \mathbb{R}^+\).
Let the grid refinement ratio be \(r := h_2 / h_1 > 1\), the experimental order of convergence be \(p \in \mathbb{R}\), and a factor of safety be \(F_s \in \mathbb{R}^+\). Then
\[ \text{GCI} := F_s \frac{\left| \frac{f_2 - f_1}{f_1} \right|}{r^p - 1} \]
is the \emph{Grid Convergence Index} (GCI)~\cite{gci}, providing an error estimate for the fine grid solution \(f_1\).
\end{definition}
To provide estimate of uncertainty, the \emph{grid convergence index} (GCI) with established~\cite{gci} safety factor \(F_s = 1.25\) was computed for the finest grid \(N=60\) as per Definition~\ref{def:gci}.
Fitting the apparent EOC for the three finest resolutions \(N \in \{20,40,60\}\) at \(p=2.1\) yields \(\text{GCI}=0.92\%\).
This low value, coupled with the fact that our finest grid result of \(\text{C}_\text{T} \simeq 0.963\) deviates from the (likely rounded) blade-resolved thrust coefficient of Ribeiro et al. by only 0.3\%, confirms that the solution is converged.

\begin{figure}
    \centering
    \includegraphics[width=\textwidth]{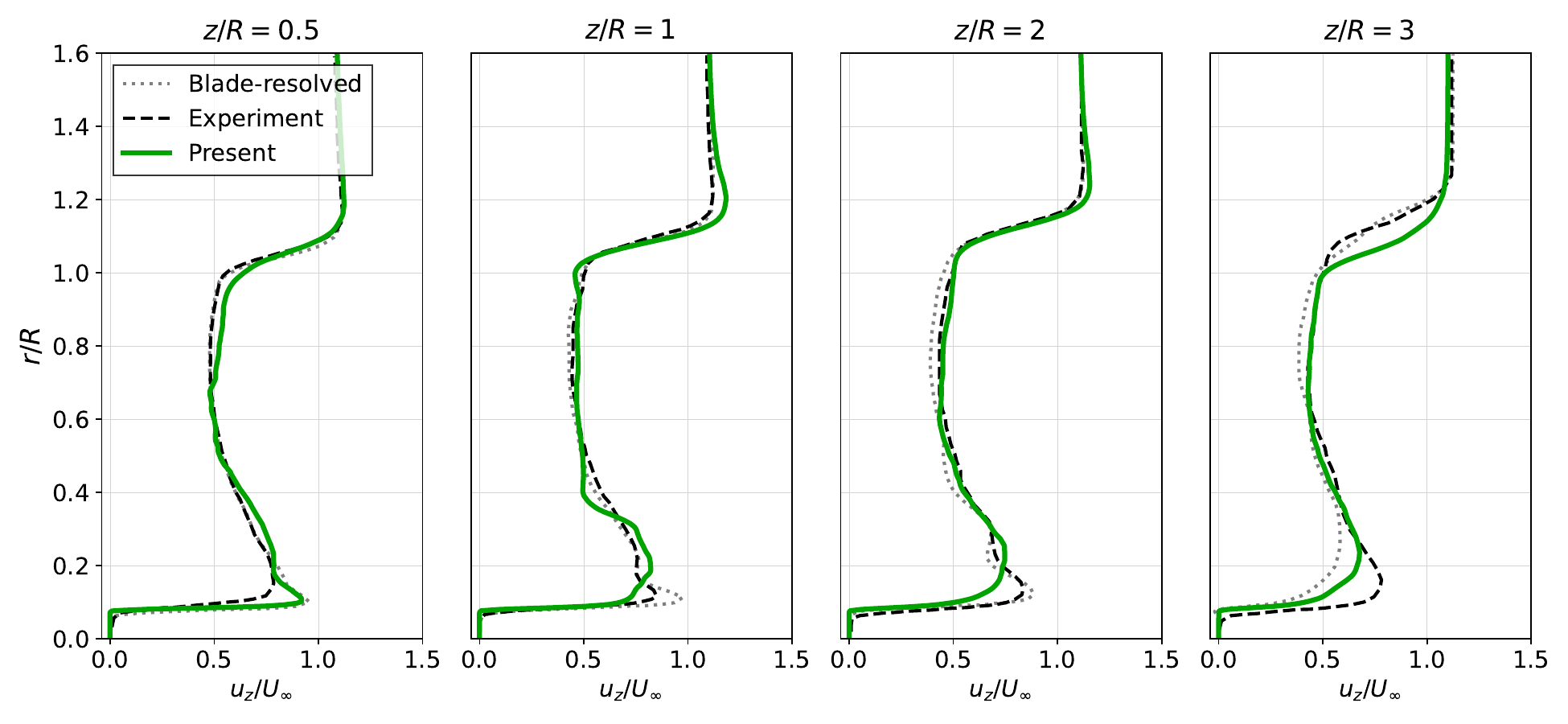}
    \caption{Time and azimuthal-averaged axial velocity at different radial lines.}
    \label{fig:lineplot}
\end{figure}

\begin{figure}
\begin{subfigure}{\textwidth}
\includegraphics[width=\textwidth]{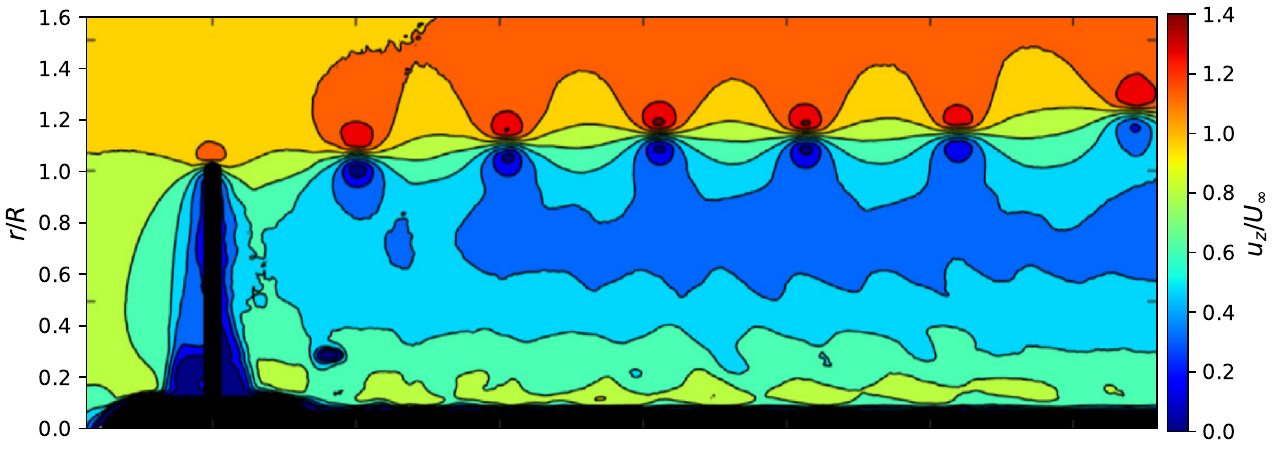}
\caption{Ribeiro et al. Experiment~\cite{ribeiroBladeresolvedActuatorLine2025}}
\end{subfigure}
\begin{subfigure}{\textwidth}
\includegraphics[width=\textwidth]{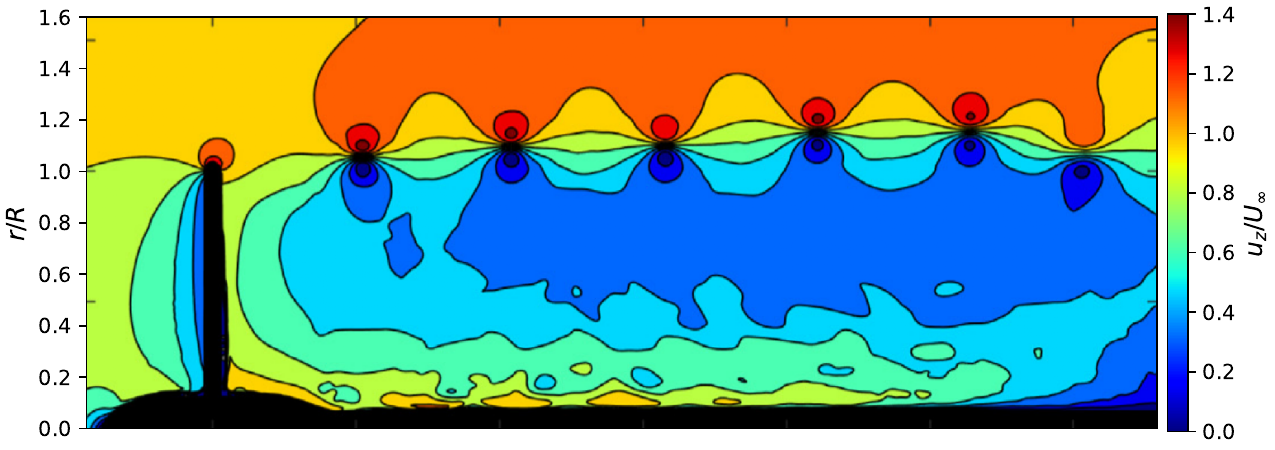}
\caption{Ribeiro et al. Blade-resolved~\cite{ribeiroBladeresolvedActuatorLine2025}}
\end{subfigure}
\begin{subfigure}{\textwidth}
\includegraphics[width=\textwidth]{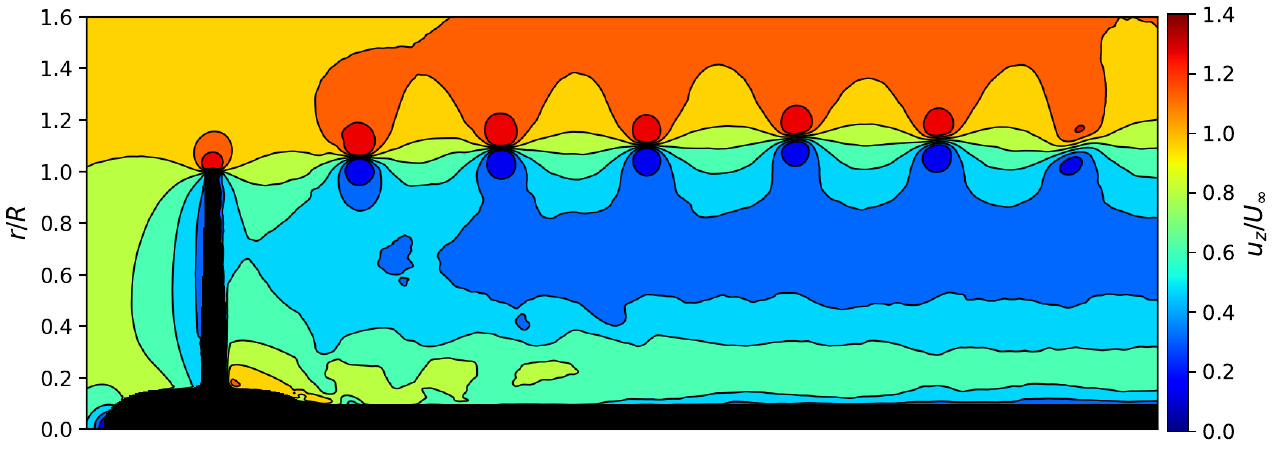}
\caption{Present}
\end{subfigure}
\caption{Comparison of phase-averaged axial velocity on a z-r plane}
\label{fig:axialU}
\end{figure}

\begin{figure}
\begin{subfigure}{\textwidth}
\includegraphics[width=\textwidth]{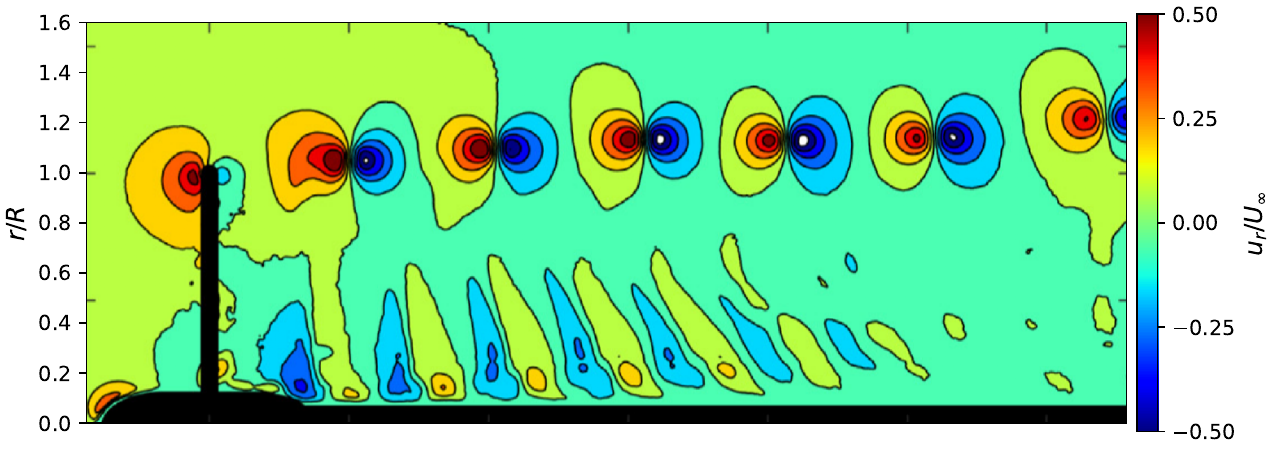}
\caption{Ribeiro et al. Experiment~\cite{ribeiroBladeresolvedActuatorLine2025}}
\end{subfigure}
\begin{subfigure}{\textwidth}
\includegraphics[width=\textwidth]{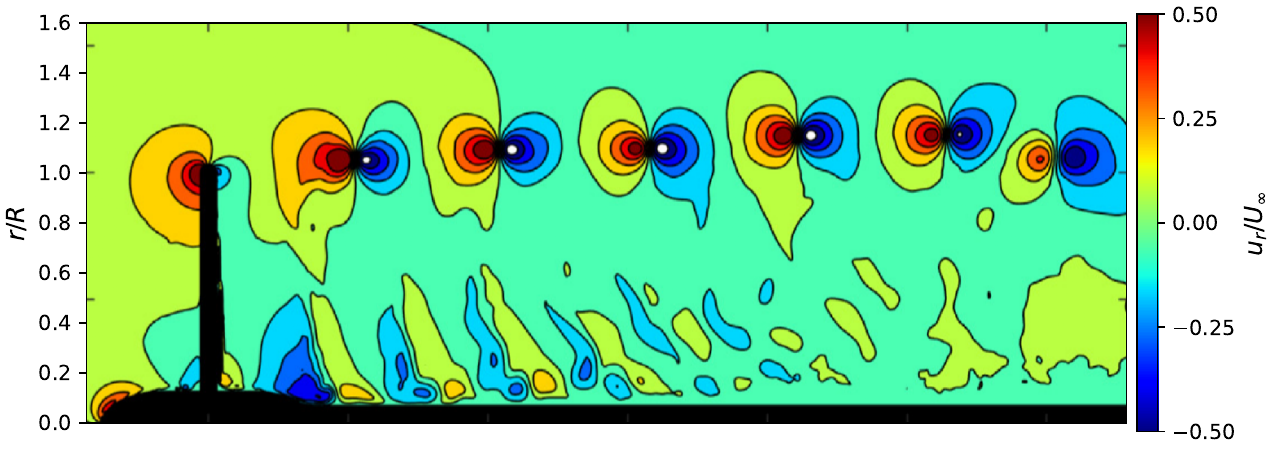}
\caption{Ribeiro et al. Blade-resolved~\cite{ribeiroBladeresolvedActuatorLine2025}}
\end{subfigure}
\begin{subfigure}{\textwidth}
\includegraphics[width=\textwidth]{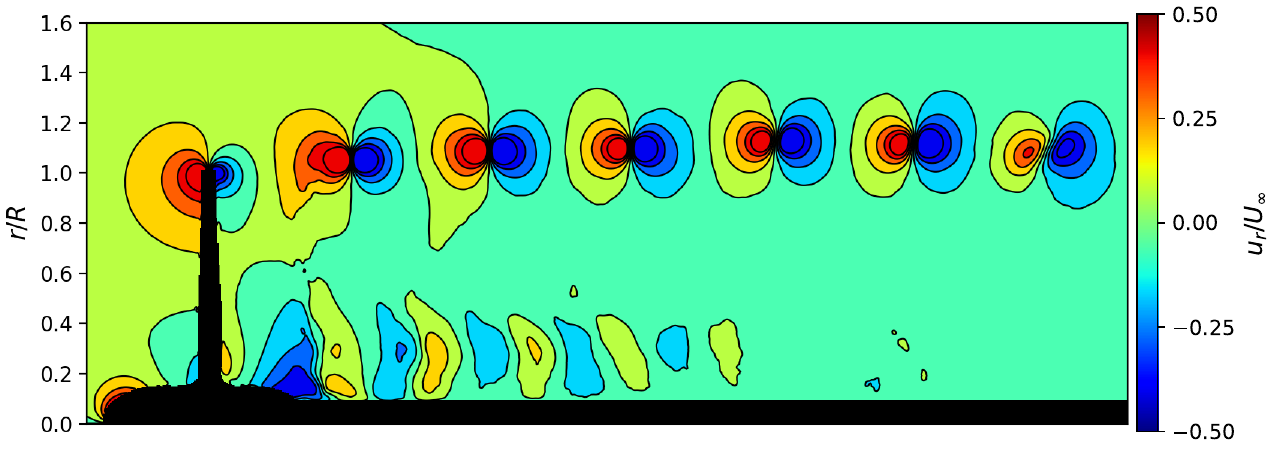}
\caption{Present}
\end{subfigure}
\caption{Comparison of phase-averaged radial velocity on a z-r plane}
\label{fig:radialU}
\end{figure}

Figure~\ref{fig:lineplot} compares the time and azimuthal-averaged axial velocity \(u_z\) along selected radial lines to the experimental and blade-resolved reference results~\cite{ribeiroBladeresolvedActuatorLine2025}.
Our present wall-modelled results compare well against both the wall-resolved LBM simulation and the experimental data. This, in addition to the matching thrust coefficient confirms that our present approach captures the dynamics of the experimental setup sufficiently well.
Figures~\ref{fig:axialU} and \ref{fig:radialU} allow for the qualitative comparison of the phase-averaged axial resp. radial velocities on a reference plane between our present approach and both the experimental as well as blade-resolved simulation results~\cite{ribeiroBladeresolvedActuatorLine2025}.

\begin{figure}
    \centering
    \includegraphics[width=\linewidth]{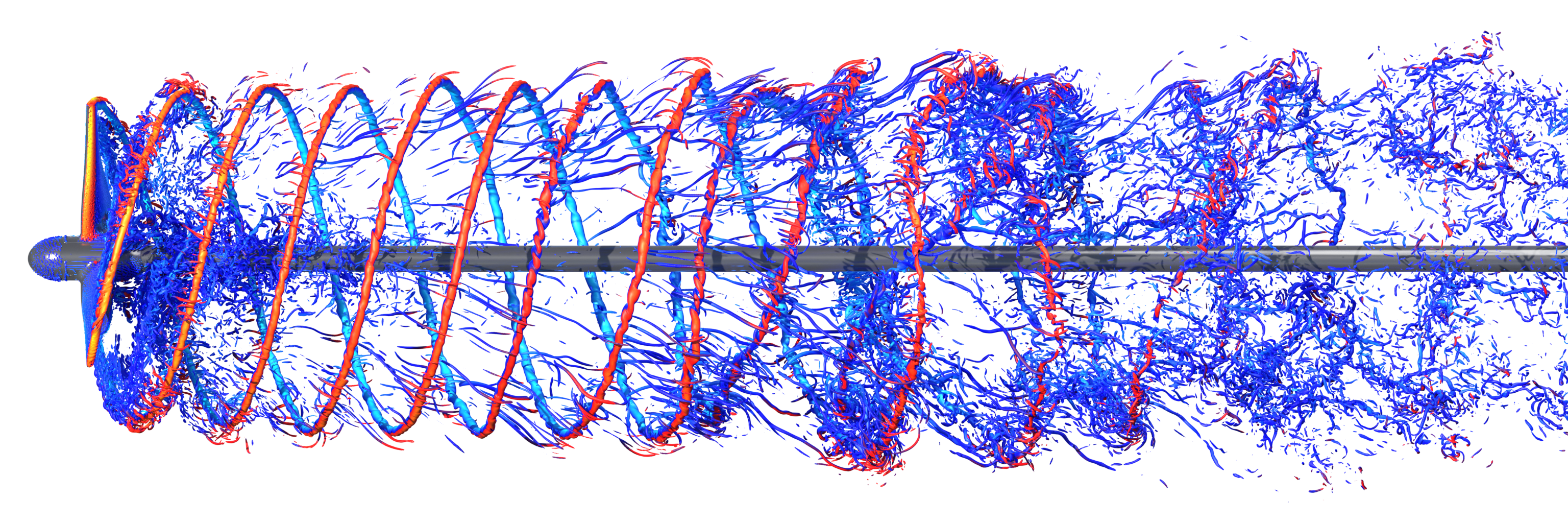}
    \caption{Rendering of Q-criterion at \(Q=\num{1e4}\) for \(N=60\), colored by axial velocity (blue toned parts move against and red tones with the axial flow direction).}
    \label{fig:qcrit}
\end{figure}

Finally, Figure~\ref{fig:qcrit} displays the contour of the Q criterion at \(Q = \qty{10000}{}\) for the simulation at \(N=60\) (equivalent to \num{4e9} cells) to showcase the characteristic tip vortices as well as their leapfrogging behavior.
The rendering was produced using Blender~\cite{blender} and the SciBlend~\cite{sciblend} extension.

\section{Performance}\label{sec:performance}

\begin{table}[h]
\centering
\begin{tabular}{l l l l l l l l l}
\toprule
Dynamics & Bandwidth [bytes] & \multicolumn{3}{l}{Operations [FLOPs]} & \multicolumn{3}{l}{Performance [MLUPs]} \\
& & No-CSE & CSE & Reduction & No-CSE & CSE & Speedup \\
\midrule
\phantom{HH}RRLBM & 152 & 4512 & 1189 & 2.90 & 2885 & 6411 & 2.22 \\
HHRRLBM & 204 & 4697 & 1618 & 3.79 & 2373 & 4252 & 1.79 \\
\bottomrule
\end{tabular}
\medskip
\caption{Arithmetic and bandwidth intensity as well as isolated speedup on A100 GPU for used local cell models}
\end{table}

\begin{figure}
    \begin{center}
    \includegraphics[width=0.7\linewidth]{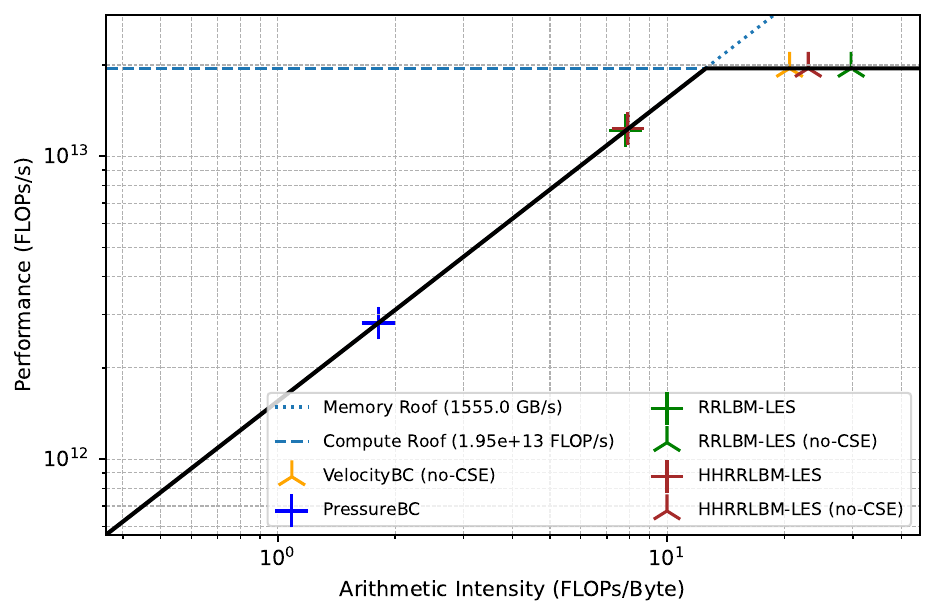}
    \end{center}
    CSE optimization is critical for reducing the arithmetic intensity of both the bulk and wall-modeled collision steps, rendering them bandwidth-limited in the ideal case. Memory and compute roofs given by theoretical maximums provided in the data sheets by NVIDIA.
    \caption{Roofline analysis of local cell models for NVIDIA A100}
    \label{fig:roofline}
\end{figure}

Due to the abstract implementation of all LB models against the \emph{concept of a cell} they are not only amenable to platform-transparent execution but also automatic code optimization using common subexpression elimination (CSE).
For the present HHRRLBM model used in the wall-modeled FSI regions, OpenLB reports a per-collision memory bandwidth of 204 bytes and a floating point complexity of 1618 FLOPs (cf. 4697 FLOPs in the unoptimized case, a \(\sim 2.9\) fold reduction).
Similarly, the bulk RRLBM model has a bandwidth of 152 bytes per cell and requires 1189 FLOPs per collision after a \(3.8\)-fold CSE optimization.
A theoretical roofline analysis for the involved dynamics utilizing OpenLB's introspection data is shown in Figure~\ref{fig:roofline}. This clearly shows that CSE is critical for moving the bulk collision steps and as such the combined problem from being compute-limited into the bandwidth-restricted region.

\begin{figure}
  \includegraphics[width=\textwidth]{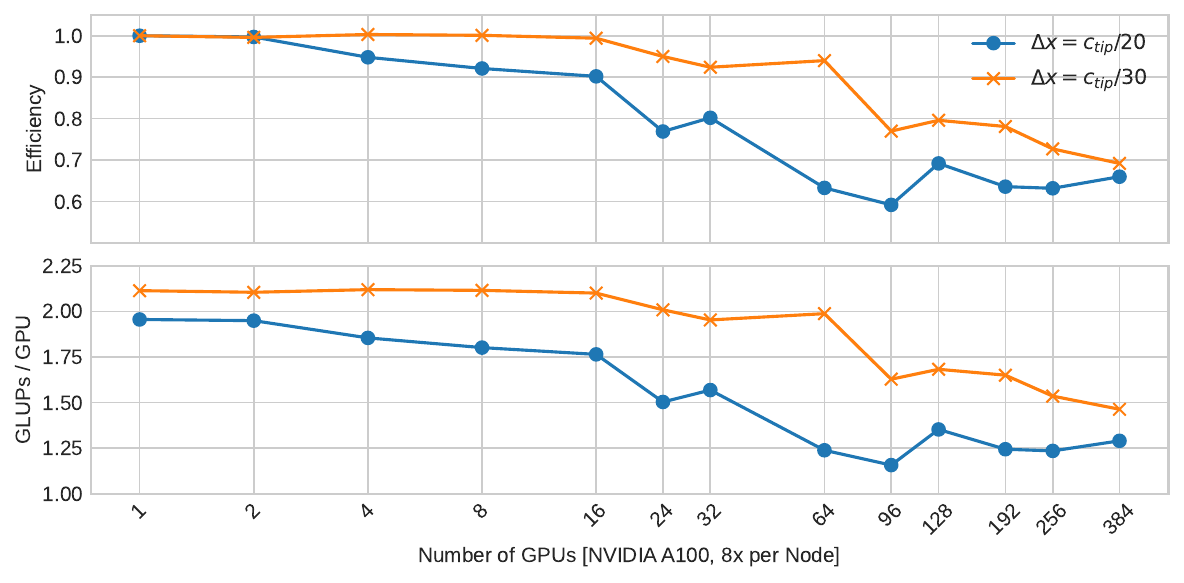}
  The weak scaling study utilizes a wind farm setup placing one turbine per A100 GPU using between 1 and 48 nodes of Karolina's accelerated partition. The setup is based directly on the present validated case, changing only the number of turbines in a \(n \times m\) grid but still performing the full wall-modeled FSI simulation with computation of the integral thrust values at every timestep.
  \caption{Weak scaling analysis of wind farm on Karolina}
  \label{fig:weakscaling}
\end{figure}

Figure~\ref{fig:weakscaling} shows the weak scaling efficiency of a exemplary wind farm case based on the validated setup (cf. Section~\ref{sec:validation}).
The study was conducted on the Karolina supercomputer, utilizing up to 384 NVIDIA A100 GPUs across 48 nodes.
Two different resolutions \(N \in \{20,30\}\) were tested, both providing high quality results w.r.t. integral reference quantities (cf. Figure~\ref{fig:eoc}). The tested problem sizes range between \num{30e6} and \num{41.02e9} cells with a peak throughput of \num{562e9} cell updates per second for this complex wall-modeled application case.
For the larger \(N=30\) case near ideal weak scaling efficiency above \qty{92}{\percent} is observed up to 64 GPUs after which the efficiency reduces until \qty{69}{\percent} for the 41 billion cell case on 384 GPUs.
This reduction also persists if all global reductions (cf. Listing~\ref{lst:hhrrlbmfsi}) are disabled, hinting at inter-node communication latency restrictions caused e.g. by non-contiguous allocation of higher node counts.

Together, these results demonstate the usability of the presented approach for large-scale blade-resolved WMLES simulations of multiple rotors on state-of-the-art GPU-accelerated supercomputers.

\clearpage
\section{Conclusion}

This paper introduced a novel framework for efficient, blade-resolved wall-modeled large eddy simulations of rotor aerodynamics. Our approach integrates a homogenized hybrid regularized recursive lattice Boltzmann method targeting the filtered Brinkman--Navier--Stokes equations with a new wall model for moving boundaries, fully implemented within the platform-transparent open-source framework OpenLB.

The method's accuracy and robustness were demonstrated through validation against experimental data and wall-resolved reference simulations of a model wind turbine.
Excellent agreement was found for integral forces, with the grid-converged thrust coefficient deviating by only \qty{0.3}{\percent} from the reference while wake velocity profiles were reproduced with high fidelity.
A grid convergence study confirmed numerical consistency, showing a clear asymptotic range with an experimental order of convergence of 1.38.

The framework's computational performance was assessed through roofline analysis and large-scale weak scaling studies on up to 384 NVIDIA A100 GPUs.
These results demonstrate high efficiency for simulations of up to 384 rotors on a 41 billion cell lattice, confirming the method’s potential for high-fidelity simulations of entire wind farms.

Nevertheless, some limitations remain. The present study focused on a single rotor geometry at moderate Reynolds number and without structural deformation.
Furthermore, the wall modeling relies on empirical functions whose accuracy -- while well established -- may be case dependent.

Clear next steps are validations against full-size wind turbines at higher Reynolds numbers, other rotor geometries and the integration of local grid refinement.
Future work can also utilize the fact that the present approach already internally represents the rotor geometry in an inertial structure lattice, to straight forwardly integrate e.g. LBM schemes targeting elastic structural deformation.
These directions aim to broaden applicability beyond wind energy to aerospace and process engineering applications where wall-modeled turbulent flows around moving structures are similarly critical.

\section*{CRediT Author Statement}

\textbf{Adrian Kummerländer:} Conceptualization, Methodology, Formal analysis, Software, Validation, Visualization, Writing - Original Draft 
\textbf{Shota Ito:} Writing - Review \& Editing
\textbf{Maximilian Schecher:} Visualization, Writing - Review \& Editing
\textbf{Davide Dapelo:}  Writing - Review \& Editing, Funding acquisition
\textbf{Stephan Simonis:} Writing - Review \& Editing, Project Management
\textbf{Mathias J. Krause:} Supervision, Writing - Review \& Editing, Resources, Funding acquisition
\textbf{Fedor Bukreev:} Supervision, Methodology, Software, Project Management, Writing - Review \& Editing

\section*{Acknowledgments}

This work has received funding from the European Union’s Horizon Europe research and innovation programme under grant agreement No 101138305. 

This work has received funding from the European Union’s Horizon Europe research and innovation programme under grant agreement No 101182847. 

This work was supported by the Ministry of Education, Youth and Sports of the Czech Republic through the e-INFRA CZ (ID:90254).

\printbibliography

\end{document}